\documentclass[twocolumn]{revtex4-1}
\begin{document}
\title{Physical Angular Momentum Separation for QED  }
\author{Weimin Sun}
\thanks{sunwm@nju.edu.cn}
\address{School of Physics, Nanjing University, Nanjing 210093, China}

\begin{abstract}
We study the non-uniqueness problem of the gauge-invariant angular momentum separation for the case of QED, which stems from the recent controversy concerning the proper definitions of the orbital angular momentum and spin operator of the individual parts of a gauge field system. For the free quantum electrodynamics without matter, we show that the basic requirement of Euclidean symmetry selects a unique physical angular momentum separation scheme from the multitude of the possible angular momentum separation schemes constructed using the various Gauge Invariant Extentions. Based on these results, we propose a set of natural angular momentum separation schemes for the case of interacting QED by invoking the formalism of asymptotic fields. Some perspectives on such a problem for the case of QCD are briefly discussed.

\end{abstract}

\maketitle

The angular momentum separation for a gauge field system (both QCD and QED) is a hot problem in recent years
and has attracted the attention of many researchers in the spin physics community. In this study, quite many
different angular momentum separation schemes for QCD (and also QED) has been proposed for different purposes.
This plurality of angular momentum separation schemes is a phenomenon existing at the present time. In this article
we shall analyze this issue from a fundamental point of view and provide a natural physical answer to this non-uniqueness problem in the case of QED, and based on these understanding, describe some possible perspectives of this problem for the case of non-abelian gauge theory.

This problem first shows its appearance in classical electromagnetism. In classical electromagnetism the angular momentum of a free electromagnetic field  has the explicit expression
\begin{eqnarray}\label{EM-AM}
{\bf J}&=&\int d^3x (E^i {\bf x}\times {\bf \nabla}A^i+{\bf E}\times {\bf A})  \nonumber \\
&=&{\bf L}+{\bf S},
\end{eqnarray}
in which a split into OAM and spin parts is clearly shown. Such a separation explicitly involves the gauge potential $A^\mu$ and is thus gauge-dependent. In the recent hot discussion on the nucleon spin structure problem \cite{LeaderLorce, Wakamatsu}, a very natural way out of this gauge non-invariance difficulty is found, that is, by invoking the concept of gauge invariant extention (GIE), one can make the apparent gauge-dependent ${\bf L}$ and ${\bf S}$ into a gauge-invariant one. This idea is physically very appealing and has been widely used in the construction of OAM and spin operators of gauge particles (gluon and photon) and Dirac fermions (quark and electron) in the context of studying internal spin structure of bound states in both QCD and QED. Unfortunately, such a natural physical idea has led to a proliferation of possible angular momentum separation schemes of both QCD and QED \cite{LeaderLorce, Wakamatsu}. Some experts in this field even believe that this plurality of angular momentum separation schemes is a natural outcome of the nontrivial interactions of the gauge field system. Physically speaking, this viewpoint is rather doubtful: if this is correct, then does it mean that the separation of the total angular momentum of a gauge field system into the contributions of its constituent fields is a dynamics-dependent (or interaction-dependent) problem, and the solution to this problem is intrinsically non-unique? In this article, we shall analyze this non-uniqueness problem for the simplest case, that is, the free abelian quantum electrodynamics without matter, and we will find that, in this simplest theory, the answer to this problem is very simple and essentially unique.

First let us recall the concept of GIE, and we shall explain the physical essence of this concept using a more direct and transparent language which differs somewhat from the one adopted in the present literature of nucleon spin structure study. The simplest and most basic example of the GIE is the Coulomb gauge one. Here, we introduce this object using a natural physical language: the Coulomb gauge GIE $A^\mu_C$ is a gauge-invariant object which gives the full field strength $F^{\mu\nu}$ just as the original $A^\mu$ does, and at the same time satisfies the condition $\partial^i A^i_C=0$, that is, this $A^\mu_C$ is the solution of the following simultaneous equations \cite{Chen2}:
\[
\left\{
\begin{array}{l}
\partial^\mu A^\nu_{C}-\partial^\nu A^\mu_{C}=F^{\mu\nu} \\
\partial^i A^i_C=0.
\end{array}
\right.
\]
It can be shown that under the natural boundary condition that $F^{\mu\nu}$ and $A^\mu_C$ vanish sufficiently rapidly at spatial infinity, the above equation has a unique solution which reads: $A^\mu_C=\frac{1}{\nabla^2}\partial^i F^{i\mu}=A^\mu+\partial^\mu \frac{{\bf \nabla}\cdot {\bf A}}{\nabla^2}$. For a free electromagnetic field, one has $A^\mu_C=(0,{\bf A}_\bot)$, which clearly shows the physical essence that $A^\mu_C$ represents $A^\mu$ in the Coulomb gauge by a gauge-invariant expression.

The above example shows the general construction of a GIE. The procedure consists in two steps. First, one chooses an arbitrary "gauge condition" $F[A]=0$, then we define the corresponding GIE $A^\mu_F$ to be a gauge-invariant object which satisfies the following two requirements:
\[
\left\{
\begin{array}{l}
\partial^\mu A^\nu_{F}-\partial^\nu A^\mu_{F}=F^{\mu\nu} \\
F[A_{F}]=0,
\end{array}
\right.
\]
Mathematically, the construction of such an $A^\mu_{F}$ is structurally equivalent to selecting a gauge potential $A^\mu$ which satisfies the given gauge condition $F[A]=0$, and any such $A^\mu_{F}$ can be regarded as representing
a given $A^\mu$ in the gauge $F[A]=0$ in a gauge-invariant manner. When the condition $F[A]=0$ uniquely fixes the gauge, the construction of $A^\mu_F$ will be unique, otherwise, one has a whole set of $A^\mu_F$ corresponding to the gauge condition $F[A]=0$.

Then, how to construct such an $A^\mu_F$? First we note that $A^\mu_F$ is a gauge invariant object by its very definition, hence it can be expressed in terms of the field strength tensor $F^{\mu\nu}$. This is actually a simple consequence of the following natural mathematical fact: if a functional $\Phi[A^\mu]$ is gauge-invariant, then it can always be expressed in terms of the field strength tensor $F^{\mu\nu}$. This mathematical fact itself can be established by the following simple reasoning. For a gauge-invariant functional $\Phi[A^\mu]$, one has $\Phi[A^\mu]=\Phi[A^\mu+\partial^\mu\theta]~~\forall \theta$. Then, for any $A^\mu$, one can introduce the accompanying Coulomb gauge GIE $A^\mu_C=A^\mu+\partial^\mu \frac{{\bf \nabla}\cdot {\bf A}}{\nabla^2}$, and one then
has $\Phi[A^\mu]=\Phi[A^\mu+\partial^\mu \frac{{\bf \nabla}\cdot {\bf A}}{\nabla^2}]=\Phi[A^\mu_C]=\Phi[\frac{1}{\nabla^2}\partial^i F^{i\mu}]$, which proves our assertion. With this fact known, one should write
$A^\mu_F(x)=f^\mu[F^{\rho\sigma};x]$. Then, how are two different GIEs related to each other? For the case of an abelian gauge theory, such a relation is very natural and simple. Suppose we are given two GIEs $A^{\mu(1)}_{\rm GIE}(x)$ and $A^{\mu(2)}_{\rm GIE}(x)$. Since they yield the same field strength $F^{\mu\nu}$, they necessarily differ by a four-gradient term, which itself should be gauge-invariant:
\begin{eqnarray}
A^{\mu(1)}_{\rm GIE}(x)- A^{\mu(2)}_{\rm GIE}(x)&=& f^{\mu(1)}[F^{\rho\sigma};x]- f^{\mu(2)}[F^{\rho\sigma};x] \nonumber \\
&=& \partial^\mu f(x).
\end{eqnarray}
With this fact being established, one can derive a general expression for any GIE corresponding to a linear gauge condition. The recipe is very simple. Suppose we try to construct a GIE satisfying some linear condition $P_\mu A^\mu_P=0$. We could first choose some specific GIE as an initial one, then construct $A^\mu_P$ by invoking the above relation and its defining condition. Here we choose the Coulomb gauge GIE as the initial one.
Then this $A^\mu_P$ necessarily differs from $A^\mu_C$ by a four-gradient term
\begin{equation}\label{GIE}
A^\mu_P=A^\mu_C+\partial^\mu f,
\end{equation}
which yields the following relation
\begin{equation}\label{GIE-Eq}
P_\mu A^\mu_P=P_\mu A_C^\mu+P\cdot \partial f=0.
\end{equation}
From this one can solve for $f$
\begin{equation}\label{}
f(x)=-\frac{1}{P\cdot \partial}P_\mu A_C^\mu(x)+f_0(x),
\end{equation}
where the inverse operator $\frac{1}{P\cdot \partial}$ is determined using some specific boundary condition,
and $f_0=f_0[{\bf E},{\bf B};x]$ is an arbitrary solution of the homogeneous equation $P\cdot \partial f_0=0$, which itself depends functionally on ${\bf E}$ and ${\bf B}$ owing to its gauge-invariance. Therefore, the GIE we seek has the following general expression:
\begin{equation}\label{GIE-form}
A^\mu_P(x)=A^\mu_C(x)-\partial^\mu \frac{1}{P\cdot \partial}P_\mu A^\mu_C(x)+\partial^\mu f_0[{\bf E},{\bf B};x],
\end{equation}
with $f_0(x)=f_0[{\bf E},{\bf B};x]$ satisfying $P\cdot \partial f_0(x)=0$.
The last term $\partial^\mu f_0$ present
in (\ref{GIE-form}) apparently reflects the residual gauge freedom in the linear gauge $P_\mu A^\mu=0$.

With such a GIE construction, one immediately obtains a new separation of the electromagnetic angular momentum
\begin{eqnarray}\label{EM-AM-F}
{\bf J}&=&\int d^3x (E^i {\bf x}\times {\bf \nabla}A^i_F+{\bf E}\times {\bf A}_F)  \nonumber \\
&=&{\bf L}_F+{\bf S}_F,
\end{eqnarray}
which has a formally gauge-invariant appearance. Since there could exist infinitely many GIEs, one could construct infinitely many separations of this kind. This is the problem of proliferation of angular momentum separation schemes in the present literature, which reflects itself in this simple abelian gauge theory of electromagnetism.

Now, let us come back to the main problem: does the plurality of the gauge-invariant angular momentum
separation scheme reflect an intrinsic arbitrariness of the physical definition of OAM and spin operator in a gauge
theory, or is this apparent non-uniqueness of angular momentum separation scheme only a mathematical artifact, which
arises due to our unconscious ignorance of some basic physical principles?  We will show that,
in a free electromagnetic field theory, the correct definition of the OAM and spin operator is uniquely determined by some basic physical considerations. However, before we turn to this issue, let us see a related problem, that is, the commutation relation of the photon spin operator, which is still a problem under discussion in the present literature. In this discussion, we will use the basic physical requirements such as the Euclidean symmetry as a guide.

First let us analyze in a general manner how to separate the angular momentum operator of a free relativistic
particle into the OAM and spin parts. Mathematically, a free relativistic particle of mass $m$ corresponds to a definite infinite-dimensional unitary representation of the $\rm{Poincar\acute{e}}$ group, with the one-particle quantum state space being the corresponding carrier space of the representation. Among the 10
generators of the $\rm{Poincar\acute{e}}$ group, the three rotation generators $J_i$ $(i=1,2,3)$ and the three spatial translation generators $P_i$ $(i=1,2,3)$ constitute the 6 generators of the Euclidean subgroup $E(3)$.
In the general discussion of the physical feature of a free relativistic particle, invariance under spatial translations and rotations (Euclidean symmetry) is a natural starting point for a physical investigation. In this article, we try to understand and analyze the angular momentum separation problem from this viewpoint. Physically, if one tries to split the rotation generator for this particle into the contributions of the OAM and spin parts: $J_i=L_i+S_i$, then the requirement of Euclidean symmetry will entail that: (1) both ${\bf L}$ and ${\bf S}$ are vector operator under spatial rotations; (2)the three components of the spin operator ${\bf S}$ necessarily commute with the translation generators: $[S_i,P_j]=0$. These two requirements stem from basic physical considerations. In order to make our discussion more transparent, we add one more requirement: the six operators $L_i$ and $S_i(i=1,2,3)$ span a closed Lie algebra among themselves:
\begin{equation}\label{SL-Commutator}
[S_i,S_j]= i c_{ijk}S_k+i d_{ijk}L_k,
\end{equation}
\begin{equation}
[L_i,L_j]= i c'_{ijk}S_k+i d'_{ijk}L_k,
\end{equation}
\begin{equation}
[S_i,L_j]= i c''_{ijk}S_k+i d''_{ijk}L_k.
\end{equation}
This condition is a mathematical one which facilitates our discussion, but it is also natural from a physical point of view. These three requirements are our basic starting point to study the angular momentum separation problem.

Now, we will show that, the above three requirements actually imply that
the three components of the spin operator span a closed Lie algebra among themselves. We demonstrate this point in the following steps.
First, from the two commutation relations $[J_i, P_j]=i \epsilon_{ijk}P_k$ and $[S_i,P_j]=0$ we have $[L_i,P_j]=i \epsilon_{ijk}P_k$. Then, substituting (\ref{SL-Commutator}) into the relation $[P_l,[S_i,S_j]]=0$ gives
\begin{equation}
[P_l, c_{ijk}S_k+d_{ijk}L_k]=0,
\end{equation}
which implies $d_{ijk}\epsilon_{klm}P_m=0$. This immediately gives $d_{ijk}\epsilon_{klm}=0$, which implies $d_{ijk}=0$. Hence, $S_i(i=1,2,3)$ span a closed Lie algebra among themselves $[S_i,S_j]= i c_{ijk}S_k$.

Then, what is the possible form of the spin operator commutation relation? We use Euclidean symmetry to find the answer. We note that ${\bf S}$ is a vector operator, hence one has the following operator transformation rule: $U(R) S_i U^{-1}(R)=(R^{-1})_{ij}S_j$, with $R$ being a rotation matrix and $U(R)$ the corresponding rotation operator. Performing the operation $U(R)\cdots U^{-1}(R)$ on both sides of the equation $[S_i,S_j]=i c_{ijk} S_k$, one has
\begin{eqnarray}
U(R)[S_i,S_j] U^{-1}(R) &=& (R^{-1})_{im} (R^{-1})_{jn} [S_m,S_n] \nonumber \\
&=& (R^{-1})_{im} (R^{-1})_{jn} i c_{mnt}S_t \nonumber \\
&=& i c_{ijk} (R^{-1})_{kt} S_t,
\end{eqnarray}
which yields a numerical relation
\begin{equation}
(R^{-1})_{im}(R^{-1})_{jn}c_{mnt}=c_{ijk}(R^{-1})_{kt}=c_{ijk} R_{tk},
\end{equation}
and from this one obtains
\begin{equation}
(R^{-1})_{im}(R^{-1})_{jn}(R^{-1})_{st} c_{mnt}=c_{ijk} (R^{-1})_{st} R_{tk} =c_{ijs}.
\end{equation}
This shows that $c_{ijk}$ should be a third-order numerical tensor under spatial rotations, hence it must be proportional to the Levi-Civita tensor
$\epsilon_{ijk}$, and consequently the spin operator commutation relation takes the form:
$[S_i,S_j]=i \kappa \epsilon_{ijk} S_k$.

The remaining problem is then to find all possible values of the constant $\kappa$ from physical considerations. To this end, let us impose one more natural requirement: ${\bf L}\cdot {\bf P}=0$, which accords with the intuitive physical picture that ${\bf L}$ represents the orbital part of the angular momentum of the particle. This immediately gives ${\bf J}\cdot {\bf P}={\bf S}\cdot {\bf P}$. Then, using the previous results, we obtain the following two commutator relations
\begin{equation}\label{CommutatorJPS}
[{\bf J}\cdot {\bf P},S_i]=[J_k P_k,S_i]=[J_k,S_i]P_k=i \epsilon_{ijk} S_j P_k,
\end{equation}
\begin{equation}\label{CommutatorSPS}
[{\bf S}\cdot {\bf P},S_i]=[S_k P_k,S_i]=[S_k,S_i]P_k=i \kappa \epsilon_{ijk} S_j P_k.
\end{equation}
The equality of the above two commutators then implies ${\bf S}\times {\bf P}=\kappa~ {\bf S}\times {\bf P}$. Hence, there are two possibilities: (1) ${\bf S} \times {\bf P} \neq 0$ and $\kappa=1$; (2) ${\bf S}\times {\bf P}=0$. In the first case, the spin operators obey the standard angular momentum algebra: $[S_i,S_j]=i \epsilon_{ijk} S_k$. In the second case, after some direct calculation, one obtains
\begin{equation}
[({\bf S}\times {\bf P})_m,S_i]=i\kappa (S_m P_i-\delta_{mi} {\bf S}\cdot {\bf P})=0,
\end{equation}
which implies $\kappa=0$, and hence the three components of the spin operator commute: $[S_i,S_j]=0$. Therefore, for a free relativistic particle, its spin operator commutation relation could have only two possible forms: $[S_i,S_j]=i \epsilon_{ijk} S_k$ or $[S_i,S_j]=0$.

After establishing this mathematical fact, let us turn to the case of the photon spin operator. Physically, experimentally observed photons only have transverse polarizations. Mathematically, the physical one-photon sector is a carrier space of an infinite-dimensional representation of the ${\rm Poincar\acute{e}}$ group, whose basis could be taken as the set of improper vectors $|{\bf k},\lambda\rangle $. For a massless photon, there is no rest frame,
and it is a well-known fact that the irreducible representation of the massless photon state is characterized by E(2) symmetry, which is a rotational symmetry with respect to the propagation direction of the photon. As was said in \cite{ZhangPak}, the only frame-independent notion of spin in the gauge theory for a massless particle is
the helicity which can be described within the framework of the little group E(2) of the Lorentz group.
In an unpublished preprint \cite{Sun} we show that on the physical one-photon sector it is not possible to define a one-photon spin operator which obeys the usual angular momentum algebra. The essential idea of the argument in \cite{Sun} is like this. If a set of one-photon spin operators obeying the commutation relation $[S_i,S_j]=i \epsilon_{ijk} S_k$ could be defined, then one can construct the raising and lowering operators $S_\pm=S_1\pm iS_2$. Let us then consider a specific state vector $|{\bf k},\lambda=+1 \rangle$ which represents a photon moving along the z-direction with helicity $+1$. Acting the lowering operator $S_{-}$ on this vector produces a new state vector $S_{-}|{\bf k},\lambda=+1 \rangle$ which can be easily shown to be nonzero. This new state vector will then represent a photon moving along the z-direction but with helicity zero, thus contradicting the fact that the physical one-photon sector only contains transverse photons. This fact, together with our previous mathematical fact, shows that a set of photon spin operators defined on the physical one-photon sector should necessarily satisfy the commutation relation $[S_i,S_j]=0$. Here, it should be mentioned that, in 1994, Van Enk and Nienhuis \cite{VanEnkNienhuis} investigated the problem of spin and OAM of photon in the framework of second quantized electromagnetic field theory in the Coulomb gauge, and obtained an explicit form of the second quantized photon spin and OAM operators which they call ${\bf S}_{rad}$ and ${\bf L}_{rad}$ (the construction of these operators will be mentioned below when we discuss the decomposition of total angular momentum operator of free quantum electromagnetic field in Coulomb gauge).
By its explicit expression these authors demonstrate that the second quantized ${\bf S}_{rad}$ operator satisfies the
commutation rule $[S_{rad~i},S_{rad~j}]=0$, whereas the usual quantum-mechanical spin operator for a
spin-one particle, which is defined as $(\hat{S}_k)_{ij}=-i\epsilon_{ijk}$ and acts on classical field modes, satisfies the standard angular momentum algebra $[\hat{S}_i,\hat{S}_j]=i \epsilon_{ijk} \hat{S}_k$. In our above analysis, using the requirement of Euclidean symmetry and some additional mathematical assumptions, we determine that the photon, as a relativistic particle, could only have two possible types of spin operator commutation relations, and furthermore, using the massless feature of the photon, we argue that the physical photon spin operator should satisfy
the commutation relation $[S_i,S_j]=0$. 

The concrete construction for such a one-photon spin operator was first proposed by the researchers in the optics community. In Ref. \cite{Bliokh}, the authors give the following explicit construction: ${\hat{\bf S}}'={\bf \kappa}({\bf \kappa} \cdot \hat{\bf S})$, with $\hat{\bf S}$ being the usual spin operator for a spin-one particle mentioned above and ${\bf \kappa}={\bf k}/|{\bf k}|$. This construction has also been described by us in \cite{Sun} starting from a natural physical definition for the photon spin: ${\bf S}_{\rm phys} |{\bf k},\lambda \rangle =\lambda \frac{{\bf k}}{|{\bf k}|} |{\bf k},\lambda \rangle$. These two constructions are in fact identical with each other, hence showing that the construction of the quantum mechanical one-photon spin operator indeed has a physically satisfactory answer.

With these facts being understood, we turn to the main problem we are to discuss, that is,
how to construct a proper OAM and spin operator for a free quantum electromagnetic field?
First, as we discuss above, in such a problem the requirement of Euclidean symmetry plays a essential role. In
a Euclidean-invariant quantum field system, the construction of the OAM and spin operator
should respect the basic requirement of Euclidean symmetry. In our case of a free quantum
electromagnetic field, we need to impose the following two requirements:
(1) the OAM and spin operator are vector operators under spatial rotations; (2) the three components of the spin
operator commute with the spatial translation generators.
These two requirements are structurally equivalent to the corresponding ones for a single free relativistic particle,
but now should be understood as an operator relation on the whole photon Fock space.
For our case of the free electromagnetic field, we should also impose one more requirement: the photon spin operator defined on the whole photon Fock space should coincide with the one-photon spin operator when restricted to
the physical one-photon sector. These three requirements are our basic premise for the discussions to
follow.

The discussion is then rather straightforward. We first note that the whole formulation of the quantized
electromagnetic field theory should respect the requirement of Euclidean symmetry. This means that the
quantization procedure should be formulated in a Euclidean-covariant gauge. Two basic quantization formalism
of the free electromagnetic field are the Coulomb gauge quantization and the Gupta-Bleuler (GB) formulation,
both of which are Euclidean covariant. The Coulomb gauge choice has the merit of only producing physical transverse photons in its Fock space, and we shall first discuss our problem in this gauge choice. When everything is clearly understood in this way, we will then turn to discuss the same problem in the GB formulation.

Now, we first see the Coulomb gauge choice. In this gauge, only the transverse component ${\bf A}_\bot$
plays a dynamical role and is quantized, the $A^0$ component simply vanishes and hence does not appear.
The angular momentum operator takes the following appearance
\begin{eqnarray}\label{Coulomb-AM}
{\bf J}&=&\int d^3x (E^i {\bf x}\times {\bf \nabla}A^i_\bot+{\bf E}\times {\bf A}_\bot)  \nonumber \\
&=&{\bf L}+{\bf S},
\end{eqnarray}
where a usual separation into OAM and spin parts is explicitly shown. Classically, such a separation
scheme is obtained using the Coulomb gauge GIE and hence it is gauge-invariant. Now, this separation
naturally satisfies the requirements of Euclidean symmetry: both ${\bf L}$ and ${\bf S}$ are vector operators, and the spin operator ${\bf S}$ commutes with the translation generator: $[S_i,P_j]=0$.
A further notable property of this separation is that both ${\bf L}$ and ${\bf S}$ are conserved.
Then, how is this separation connected with the OAM-spin separation of the one-photon angular momentum
operator? To see this, we use the plane wave expansion of the ${\bf A}_\bot$ operator to make the calculation
and find
\begin{equation}\label{Coulomb-S}
{\bf S}=\int \frac{d^3k}{(2\pi)^3} \sum_{\lambda=\pm1} \lambda \frac{{\bf k}}{|{\bf k}|}a_{{\bf k},\lambda}^\dagger a_{{\bf k},\lambda},
\end{equation}
with $a_{{\bf k},\lambda}^\dagger$ and $a_{{\bf k},\lambda}$ being the photon creation and annihilation operators. Such an ${\bf S}$ is an operator acting on the whole photon Fock space $\mathcal{H}=\bigoplus\limits_{n=0}^\infty \mathcal{H}_n$, and has the following explicit action on the many-photon states:
\begin{eqnarray}\label{Spin-action}
{\bf  S}|0\rangle&=& 0 \nonumber \\
{\bf  S}a_{{\bf k},\lambda}^\dagger|0\rangle &=& \lambda \frac{{\bf k}}{|{\bf k}|}a_{{\bf k},\lambda}^\dagger|0\rangle  \nonumber \\
\cdots \cdots \nonumber \\
{\bf  S}a_{{\bf k}_1,\lambda_1}^\dagger \cdots a_{{\bf k}_n,\lambda_n}^\dagger|0\rangle &=& \bigg( \sum_{i=1}^n \lambda_i \frac{{\bf k}_i}{|{\bf k}_i|} \bigg) a_{{\bf k}_1,\lambda_1}^\dagger \cdots a_{{\bf k}_n,\lambda_n}^\dagger|0\rangle
\end{eqnarray}
Thus, the operator ${\bf S}$ coincides with the one-photon spin operator ${\bf S}_{\rm phys}$ on the one-photon sector: ${\bf S}_{\rm phys}={\bf S}|_{\mathcal{H}_1}$, and it obeys the commutation rule $[S_i,S_j]=0$,
since each component of this operator is a superposition of particle-number operators. In this sense, the operator ${\bf S}$ appearing in the separation (\ref{Coulomb-AM}) is a natural extension of the one-photon spin operator
${\bf S}_{\rm phys}$ to the whole photon Fock space. Therefore, we can say that the separation (\ref{Coulomb-AM}) provides a physically satisfactory definition of the OAM and spin operator for the free electromagnetic field.

Then, how about other angular momentum separations obtained using a different GIE construction?  First, we note that, similar to its classical counterpart, the quantum angular momentum operator (\ref{Coulomb-AM}) is kept invariant under an arbitrary operator-valued gauge transformation: $A^\mu_C \rightarrow  A^\mu_C+\partial^\mu f$, hence using any specific (operator-valued) GIE $A^\mu_F$ we can construct a new angular momentum separation:
\begin{eqnarray}\label{GIE-AM}
{\bf J}&=&\int d^3x (E^i {\bf x}\times {\bf \nabla}A^i_F+{\bf E}\times {\bf A}_F)  \nonumber \\
&=&{\bf L}_F+{\bf S}_F.
\end{eqnarray}
Thus, we could construct infinitely many angular momentum operator separations, all of which are gauge-invariant.
Then, can all these angular momentum separation schemes be physically acceptable? The crucial point is the Euclidean symmetry, which we shall discuss in detail below.

First, we shall restrict ourselves to considering GIEs defined by a linear condition. This is mainly for the purpose of simpicity. Suppose we look at a GIE $A^\mu_P$ satisfying some linear condition $P_\mu A^\mu_P(x)=0$. Then, Euclidean symmetry requires that the condition $P_\mu A^\mu_P(x)=0$ be a Euclidean invariant one. In fact, the spin operator ${\bf S}_P$ has the form of a volume integral of a local density: ${\bf S}_P=\int d^3x {\bf E}\times {\bf A}_P$, hence the global requirements imposed on the ${\bf S}_P$ operator entail that ${\bf A}_P(x)$ be a three-vector field and transform in the same way as ${\bf A}_\bot(x)$ does under spatial translations.  The GIE operator ${\bf A}_P(x)$ can be constructed in the following way:
\begin{equation}\label{GIE-P-form}
A^i_P(x)=A^i_\bot(x)+\partial^i \frac{1}{P^0 \partial^0+P^i\nabla^i}P^i A^i_\bot(x)+\partial^i f_0[{\bf E},{\bf B};x],
\end{equation}
where $f_0(x)$ is an operator-valued function satisfying $P\cdot \partial f_0(x)=0$. Hence, the global structure of
the above construction shows that the GIE condition $P_\mu A^\mu_P(x)=0$ should be Euclidean invariant, i.e.,
the $P^0$ component be a rotational scalar, the $P^i$ components form a three-vector, and all the four components of $P^\mu$ have no explicit $x^\mu$ dependence, so that the above requirements on the GIE operator ${\bf A}_P(x)$
could be satisfied.

The Coulomb condition ${\bf \nabla}\cdot {\bf A}_C=0$ meets this requirement, but it only involves the spatial vector
potential. Generally speaking, a GIE condition should also contain the $A^0$ component. If we assume the GIE
condition $P_\mu A^\mu_P(x)=0$ to be Euclidean invariant and do not contain differential operations higher than first
order, then all the possible forms of this GIE condition can be listed below:
(1)~$A^0=0$,~(2)~$\frac{\partial }{\partial t}A^0+\kappa {\bf \nabla}\cdot {\bf A}=0$~($\kappa$ is a dimensionless constant),~(3) $A^0+\frac{1}{M}{\bf \nabla}\cdot {\bf A}=0$~($M$ is an arbitrary mass scale).
The first form defines the so-called temporal gauge GIE, which will be examined closely below. The second form is structurally similar to the usual Lorenz condition, and will be examined when we gain enough insight from the discussion of the temporal gauge GIE case. The third form looks somewhat peculiar, and there is an arbitrary mass scale contained in it. As will be seen in the discussion below, it is physically unnatural to permit an arbitrary
mass scale to enter into the definition of the basic observables of the system, such as ${\bf L}$ and ${\bf S}$
operator of a free electromagnetic field, hence it is unnecessary to consider such a GIE condition.
Here, we also remark that in the discussion of the photon spin decomposition problem, other popular GIEs, such as those based on the light-cone (LC) gauge $A^{+}=0$ and the spatial axial gauge $A^{3}=0$, are also used. In this article, we restrict ourselves to considering those GIE conditions that satisfy the E(3) symmetry, and hence only consider the above-mentioned GIE conditions.

Now, we first consider the temporal gauge GIE construction.
Using the expression (\ref{GIE-form}) one can obtain the following
general form of the temporal gauge GIE:
\begin{equation}
A^\mu_T(x) = A^\mu_C(x)+\partial^\mu f_0[{\bf E},{\bf B};x],
\end{equation}
where $f_0$ is an arbitrary gauge-invariant object satisfying $\frac{\partial }{\partial t}f_0[{\bf E},{\bf B};x]=0$. With such an $A^\mu_T$ operator, one can obtain a new angular momentum operator separation
\begin{eqnarray}\label{temporal-AM}
{\bf J}&=&\int d^3x (E^i {\bf x}\times {\bf \nabla}A^i_T+{\bf E}\times {\bf A}_T)  \nonumber \\
&=&{\bf L}_T+{\bf S}_T.
\end{eqnarray}
Therefore, we can have a class of gauge-invariant separation schemes, the Coulomb gauge one being just a special
member of this class. Now, we argue that any separation scheme in this class that differs from the Coulomb gauge one is physically unacceptable because some basic physical requirements cannot be satisfied. We will explain this issue from the following three aspects.

The first point is about fundamental physical understanding. As we see above, in the Coulomb gauge separation (\ref{Coulomb-AM}) both ${\bf L}$ and ${\bf S}$ are conserved. This nice property is an inherent feature of
Coulomb gauge separation and should be preserved in any other possible angular momentum separation scheme. Now, if we check this point for a general temporal gauge separation scheme (\ref{temporal-AM}),
we will find that this nice property is lost. In fact, using the Maxwell equation for the ${\bf A}_\bot$ operator,
we obtain (ignoring any operator ordering problem at the moment)
\begin{equation}
\frac{d {\bf S}_T}{dt}=\int d^3x \nabla^2 f_0 \cdot {\bf B},
\end{equation}
which clearly demonstrates our statement.

The second point concerns the concrete physical construction. We first note that the Euclidean transformation property
of the ${\bf A}_T$ operator entails the corresponding transformation rules for $f_0[{\bf E},{\bf B};x]$: $f_0$
is a rotational scalar and transforms properly under spatial translations. These conditions put rather strong constraints on the construction of $f_0$. In the simplest case, if one assumes $f_0[{\bf E},{\bf B};x]$ to be linear in ${\bf E}$ and ${\bf B}$, then $f_0$ should be constructed using the following two expressions or be a linear superposition of them:
\begin{eqnarray}\label{}
f_0(x)&=&\int d^3y (x-y)^i f(|{\bf x }-{\bf y}|)E^i({\bf y},t_0) \\
f_0(x)&=&\int d^3y (x-y)^i f(|{\bf x }-{\bf y}|)B^i({\bf y},t_0),
\end{eqnarray}
where $f(r)$ is an arbitrary numerical function and $t_0$ some specific instant of time. However, it can be shown
that the above two expressions actually vanish due to the transverse character of ${\bf E}$ and ${\bf B}$
(this is apparent from the rotational invariance of $f(r)$ and the convolution structure of these two expressions). Physically this is just a reflection of the simple fact that one cannot construct a rotational scalar (which also transforms properly under spatial translations) using one single three-vector field ${\bf E}$ or ${\bf B}$.
Therefore, $f_0[{\bf E},{\bf B};x]$ should be at least quadratic in ${\bf E}$ and ${\bf B}$. We will argue that in this case an arbitrary mass scale (or mass parameter) necessarily enters into the expression of $f_0$, and hence is introduced into the definition of the fundamental observables ${\bf L}$ and ${\bf S}$.
Here, we will use a simple example to explain this point. In this example, we suppose $f_0$ only contains terms quadratic in ${\bf E}$ and ${\bf B}$, then all possible forms of $f_0$ can be written down using the symmetry
property constraints. To simplify our discussion, we further assume that $f_0$ is constructed in terms of the following three local rotational scalars: ${\bf E}^2$, ${\bf B}^2$ and ${\bf E}\cdot {\bf B}$. Since the last object
${\bf E}\cdot {\bf B}$ is a pseudoscalar, it is actually not allowable in the construction of $f_0$, which should
yield a polar vector field ${\bf A}_T(x)$ to ensure the axial-vector nature of the resulting spin operator. In the
simplest case, $f_0$ can be a local field, for instance, $f_0(x) \sim {\bf E}^2({\bf x},t_0)$, while in
the general case, it can have the following two possible forms or be a linear superposition of them:
\begin{eqnarray}
f_0(x)=\int d^3y  f(|{\bf x }-{\bf y}|) {\bf E}^2 ({\bf y},t_0) \label{f0-form1} \\
f_0(x)=\int d^3y  f(|{\bf x }-{\bf y}|) {\bf B}^2 ({\bf y},t_0), \label{f0-form2}
\end{eqnarray}
where $f(r)$ is some suitable numerical function. Now, $f_0$ should be a dimensionless object to ensure that ${\bf A}_T$ have mass dimension one, as required by
the natural meaning of a GIE. In the simplest case, if one assumes, for instance, $f_0(x) \sim {\bf E}^2({\bf x},t_0)$, then due to dimensional reasons one should actually write $f_0(x) \sim \frac{{\bf E}^2({\bf x},t_0)}{M^4}$, thus an arbitrary mass scale $M$ enters into the expression of $f_0$. When $f_0$ has a more general form given by the integral expressions (\ref{f0-form1}) and (\ref{f0-form2}), one can also show that the same thing happens. The demonstration is actually quite simple. In fact, if we look at our whole construction from the point of view of classical electromagnetism, then everything can be clearly seen. In classical electromagnetism,
the total field energy of the system is required to be finite:
\begin{equation}\label{}
\int d^3x \frac{1}{2}({\bf E}^2+{\bf B}^2)<+\infty,
\end{equation}
which entails the following asymptotic behaviours for ${\bf E}^2$ and ${\bf B}^2$: ${\bf E}^2({\bf x})\sim \frac{1}{r^{3+\epsilon}}$, ${\bf B}^2({\bf x})\sim \frac{1}{r^{3+\epsilon}}$ as $r \rightarrow \infty$. Now,
let us assume the following asymptotic behaviour for $f(r)$: $f(r) \sim \frac{1}{r^\alpha}$ as $r \rightarrow \infty$.
Then the convergence of the two integrals in (\ref{f0-form1}) and (\ref{f0-form2}) entails $\alpha \geq 0$.
In this case, the overall length dimension of the integral expressions (\ref{f0-form1}) and (\ref{f0-form2}), which is $\frac{1}{[L]^{1+\alpha}}$, must be nonzero, hence some mass scale must be introduced into the construction of $f_0$. The above analysis is quite typical, and one can be convinced that such a conclusion holds in the general case. Thus, in a general temporal gauge separation (\ref{temporal-AM}), it is inevitable that a mass scale has been introduced into the definition of ${\bf L}$ and ${\bf S}$. However, this is absolutely unnatural from a physical point of view, because the fundamental Lagrangian density of a free electromagnetic field contains no free mass parameters, and as a consequence, the definition of the fundamental observables of the system, such as the OAM and spin operator, should
not involve any extrinsic mass parameter whose value could be adjusted at will, otherwise, the whole formalism would
be totally arbitrary. Therefore, a temporal gauge separation scheme (\ref{temporal-AM}) cannot provide a physically correct definition of the OAM and spin operator for a free electromagnetic field, unless it coincides with the Coulomb gauge one.

The third point is actually quite simple but very persuasive. If the ${\bf A}_T$ operator contains quadratic or higher terms in ${\bf E}$ and ${\bf B}$ (or even it is some highly nonlinear expression involving ${\bf E}$ and ${\bf B}$), then the resulting spin operator ${\bf S}_T=\int d^3x {\bf E}\times {\bf A}_T$ will generally contain terms that cannot leave the one-particle sector invariant. In this case, the restriction of ${\bf S}_T$ on the one-photon sector will no longer coincide with the one-photon spin operator ${\bf S}_{\rm phys}$ and
thus completely loses its natural physical explanation as the spin operator for free photons. The above analysis
shows that in the whole class of temporal gauge separations (\ref{temporal-AM}), only the Coulomb gauge one with ${\bf A}_T={\bf A}_\bot$ can serve to provide a physically acceptable angular momentum separation scheme.

With the above discussion, we reach the conclusion that among the whole class of temporal gauge separation
schemes, only the Coulomb gauge one is the physically satisfactory choice. Then, let us see the second GIE
form defined by the condition $\frac{\partial }{\partial t}A^0+\kappa {\bf \nabla}\cdot {\bf A}=0$. In this
case, the GIE operator has the following general form:
\begin{equation}
A^\mu_{\rm GIE}(x)= A^\mu_C(x)+\partial^\mu f_0[{\bf E},{\bf B};x],
\end{equation}
with $f_0[{\bf E},{\bf B};x]$ satisfying $(\frac{\partial^2}{\partial t^2}-\kappa \nabla^2 )f_0=0$.
In this case, the Euclidean transformation property of the
operator ${\bf A}_{\rm GIE}$ will entail the corresponding transformation rule for the $f_0$ operator, which
restricts the possible forms of $f_0$. Similar to our previous discussion, one can find that $f_0$ should be at least quadratic in ${\bf E}$ and ${\bf B}$ and the whole analysis we made for the temporal gauge case also applies here. Then, the problems inherent in the temporal gauge separation schemes also exist in this case, and we conclude that a generic GIE operator $A^\mu_{\rm GIE}$ cannot yield a physically acceptable angular momentum separation scheme for a free electromagnetic field. Therefore, we finally conclude that the Coulomb gauge separation (\ref{Coulomb-AM})
is the only physically satisfactory angular momentum separation scheme in the Coulomb gauge quantization.

With the angular momentum separation problem in the Coulomb gauge quantization completely solved,
we now come to the same problem in the GB formulation. In the GB formulation, the Lagrangian density of the system
is modified to be
\begin{equation}
 \mathcal{L}=-\frac{1}{4}F_{\mu\nu}F^{\mu\nu}-\frac{1}{2}(\partial_\mu A^\mu)^2
\rightarrow -\frac{1}{2}\partial_\mu A_\nu \partial^\mu A^\nu.
\end{equation}
As is well-known, with this modified Lagrangian density all the four components of $A^\mu$ are treated as independent dynamical variables and the resultant Fock space also contains unphysical longitudinal and scalar photons. The total angular momentum operator of the system has the following explicit form
\begin{eqnarray}\label{GB-AM}
{\bf J} &=& \int d^3x \bigg(\frac{\partial A_\mu}{\partial t}{\bf x}\times {\bf \nabla}A^\mu-\frac{\partial {\bf A}}{\partial t}\times {\bf A}\bigg) \nonumber \\
  &=& {\bf L}+{\bf S}£¬
\end{eqnarray}
where a split into OAM and spin contributions is clearly shown. It can be shown that both ${\bf L}$ and ${\bf S}$ are conserved. Using the plane wave expansion of the $A^\mu$ operator, one can find
\begin{eqnarray}
{\bf S}&=&i\int \frac{d^3 k}{(2\pi)^3}\Big\{ \frac{{\bf k}}{|{\bf k}|}(a^\dagger_{{\bf k},2} a^{}_{{\bf k},1}-a^\dagger_{{\bf k},1}a^{}_{{\bf k},2})+{\bf \epsilon}^{(1)}({\bf k})(a^\dagger_{{\bf k},3}a^{}_{{\bf k},2} \nonumber \\
&&-a^\dagger_{{\bf k},2}a^{}_{{\bf k},3})+{\bf \epsilon}^{(2)}({\bf k})(a^\dagger_{{\bf k},1}a^{}_{{\bf k},3}-a^\dagger_{{\bf k},3}a^{}_{{\bf k},1}) \Big\},
\end{eqnarray}
where $\big({\bf \epsilon}^{(1)}({\bf k}),{\bf \epsilon}^{(2)}({\bf k}),\frac{{\bf k}}{|{\bf k}|}\big)$ is a right-handed dreibein of spatial polarization vectors. This spin operator contains the contribution of three types of quanta: two kinds of transverse photons and the longitudinal photon, which all have spin one (the scalar photon has spin zero and thus should not contribute). It is easy to check that, as an operator acting on the whole photon Fock space, the spin operator ${\bf S}$ obeys the standard angular momentum algebra: $[S^i,S^j]=i \epsilon^{ijk}S^k$. Now let us look at the one-photon sector $\mathcal{H}_{1{\rm GB}}$ which accommodates these four types of photons. The spin operator ${\bf S}$ leaves this one-photon sector invariant and has the explicit action:
\begin{eqnarray}
{\bf S}~a^\dagger_{{\bf k},0}|0\rangle &=& 0 \nonumber \\
{\bf S}~a^\dagger_{{\bf k},1}|0\rangle &=& i\Big( +\frac{{\bf k}}{|{\bf k}|} a^\dagger_{{\bf k},2}|0\rangle-{\bf \epsilon}^{(2)}({\bf k})a^\dagger_{{\bf k},3}|0\rangle \Big) \nonumber  \\
{\bf S}~a^\dagger_{{\bf k},2}|0\rangle &=& i\Big( -\frac{{\bf k}}{|{\bf k}|} a^\dagger_{{\bf k},1}|0\rangle+{\bf \epsilon}^{(1)}({\bf k})a^\dagger_{{\bf k},3}|0\rangle \Big)  \nonumber \\
{\bf S}~a^\dagger_{{\bf k},3}|0\rangle &=& i\Big( -{\bf \epsilon}^{(1)}({\bf k})a^\dagger_{{\bf k},2}|0\rangle+{\bf \epsilon}^{(2)}({\bf k})a^\dagger_{{\bf k},1}|0\rangle \Big).
\end{eqnarray}
Because of this, one naturally has: $[S^i|_{\mathcal{H}_{1{\rm GB}}},S^j|_{\mathcal{H}_{1{\rm GB}}}]=i\epsilon^{ijk}S^k|_{\mathcal{H}_{1{\rm GB}}}$. Now, the true observable photons are those with transverse polarizations and the corresponding physical subspace $\mathcal{H}_{1\bot} \subset \mathcal{H}_{1{\rm GB}}$ is
spanned by the vectors $a^\dagger_{{\bf k},1}|0\rangle$ and $a^\dagger_{{\bf k},2}|0\rangle$. If one only concentrates on transverse photons, one can introduce the projection of the one-photon spin operator ${\bf S}|_{\mathcal{H}_{1{\rm GB}}}$ on the physical subspace $\mathcal{H}_{1\bot}$:
${\bf S}_{\rm tr}=P_\bot{\bf S}|_{\mathcal{H}_{1{\rm GB}}} P_\bot$, where $P_\bot$ is the projection operator corresponding to the physical subspace $\mathcal{H}_{1\bot}$. The action of ${\bf S}_{\rm tr}$ on the whole one-photon
sector $\mathcal{H}_{1{\rm GB}}$ is given as:
\begin{eqnarray}
{\bf S}_{\rm tr}a^\dagger_{{\bf k},0}|0\rangle &=& 0 \nonumber \\
{\bf S}_{\rm tr}a^\dagger_{{\bf k},1}|0\rangle &=& i \frac{{\bf k}}{|{\bf k}|} a^\dagger_{{\bf k},2}|0\rangle \nonumber  \\
{\bf S}_{\rm tr}a^\dagger_{{\bf k},2}|0\rangle &=& -i \frac{{\bf k}}{|{\bf k}|} a^\dagger_{{\bf k},1}|0\rangle  \nonumber \\
{\bf S}_{\rm tr}a^\dagger_{{\bf k},3}|0\rangle &=& 0.
\end{eqnarray}
If one introduces the usual photon states with circular polarizations: $a^\dagger_{{\bf k},\lambda=\pm1}|0\rangle
=\frac{a^\dagger_{{\bf k},1}\pm i a^\dagger_{{\bf k},2}}{\sqrt{2}}|0\rangle$, then one has
\begin{eqnarray}
{\bf S}_{\rm tr}a^\dagger_{{\bf k},\lambda=\pm1}|0\rangle = \lambda \frac{{\bf k}}{|{\bf k}|}a^\dagger_{{\bf k},\lambda=\pm1}|0\rangle,
\end{eqnarray}
which agrees exactly with the action of the one-photon spin operator ${\bf S}_{\rm phys}$, and on the physical subspace $\mathcal{H}_{1\bot}$, one has $[S^i_{\rm tr}|_{\mathcal{H}_{1\bot}},S^j_{\rm tr}|_{\mathcal{H}_{1\bot}}]=0$. Therefore, we conclude that the angular momentum separation (\ref{GB-AM}) has a natural correspondence with the previous one in the Coulomb gauge quantization.

Now, one will ask a natural question: how does the non-uniqueness problem of the gauge-invariant angular momentum
separation show itself in the case of GB formulation? As to this issue, we first note that, in the GB formulation, the gauge potential $A^\mu(x)$ contains unphysical degrees of freedom and hence cannot be expressed in terms of the field strength tensor $F^{\mu\nu}$.  In fact, in the GB formulation, it can be shown that \cite{Strocchi}, an operator is gauge-invariant (in the sense of gauge transformation of the c-number type: $A^\mu \rightarrow A^\mu+\partial^\mu f$, $\partial_\mu\partial^\mu f=0$) if and only if its action on the vacuum state produces a physical state. Therefore, in the GB formulation the operator $A^\mu(x)$ is not an operator-valued GIE. The second point is that in the GB formulation the total angular momentum vector is not gauge-invariant, even not invariant under the restricted c-number gauge transformation $A^\mu \longrightarrow A^\mu+\partial^\mu f$, $\partial_\mu\partial^\mu f=0$, hence the degree of freedom of performing an operator-valued gauge transformation $A^\mu \longrightarrow A^\mu+\partial^\mu f$ simply does not exist for the total angular momentum operator. Because of these two reasons, it is completely unnecessary to
try to seek a different angular momentum separation scheme using the usual construction of GIE in the GB formulation,
and the angular momentum separation scheme (\ref{GB-AM}) is a perfect one.

From all the above discussion, one may conclude that for a free quantum electrodynamics without matter, the physically
correct angular momentum separation scheme, in other words, the correct definition of the OAM and spin operator, is
essentially unique, and is determined by the basic physical considerations.

Here, we also would give some further discussion on this issue. We have shown that the GIE based on the Coulomb gauge is the only consistent one in the decomposition of the free photon angular momentum. In fact, this reflects the uniqueness of the so-called Helmholz decomposition, also called the transverse-longitudinal
decomposition. In the standard textbook of electromagnetism, it is stated that the total angular momentum of the photon cannot be decomposed into its intrinsic spin and orbital parts so as to satisfy the gauge-invariance and the
Lorentz-invariance at the same time. One of the important features of the familiar photon spin decomposition is that one can discard the second requirement, i.e. the Lorentz-invariance. This is because the measurements
of the photon spin and OAM are carried out in a prescribed Lorentz frame, i.e. in the laboratory system.
Once the Lorentz frame is fixed, the total photon field can be decomposed into the transverse and longitudinal parts uniquely. This is assured by the uniqueness theorem due to Helmholz. This transverse-longitudinal decomposition is nothing but the GIE based on the Coulomb gauge. For the free photon case (i.e., in the absence of the charged particle sources), one can set both of $A^0$ and $A^3$ (the scalar and longitudinal components of the photon) to be zero
simultaneously by using the gauge transformation. Namely, in this case, one is left with only the two physical degrees of freedom $A^1_\perp$ and $A^2_\perp$ with $A^0=A^3=0$ (such a choice is called the "helicity gauge" in Ref.\cite{ZhangPak}). This means that the Helmholz decomposition in fact exhausts all the GIEs.
This fact can also be interpreted in another way. Under the usual condition that $A^\mu$
tend to zero sufficiently rapidly at spatial infinity, there is no residual gauge
degrees of freedom in the Coulomb gauge, this also demonstrates the Helmholz decomposition theorem, i.e. the uniqueness of the transverse-longitudinal decomposition. Then, if one further confines to the free
photon case, one can set $A^0 = 0$ since $A^0$ satisfies Laplace equation $\Delta A^0 = 0$ in this free case
(the longitudinal component $A^3$ can always be set zero by using  gauge transformation, irrespectively of the presence or absence of the charged particle sources.). In this way, all these gauges, such as the temporal gauge,
the spatial axial gauge and the LC gauge, reduce to the Coulomb gauge case. This also demonstrates the conclusion we said above.

Now, the problem of the definition of the OAM and spin operator for a free electromagnetic field has a definite solution. How about the angular momentum separation for an interacting QED theory? For simplicity, we assume the
electromagnetic field is coupled with a single flavor of Dirac fermion, which we call electron. In this interacting
field theory, a formal solution to the angular momentum separation problem always exists by invoking the formalism
of asymptotic fields. The idea is very simple. According to the the usual assumption of asymptotic completeness, the free-particle Fock space of in-states and out-states coincides with the Hilbert space of the interacting field theory: $\mathcal{H}_{\rm in}=\mathcal{H}_{\rm out}=\mathcal{H}$, hence, if one looks at the angular momentum separation problem on the free-particle Fock space, a natural solution can be found, because the structure of the free-particle
Fock space $\mathcal{H}_{\rm in}$ and $\mathcal{H}_{\rm out}$ is identical to that of a free field theory. In the subsequent discussion, we choose the Coulomb gauge quantization. In the Coulomb gauge quantization, the angular momentum operator for the free particle (photon and electron) system (both in-states and out-states) has the following
explicit expressions:
\begin{eqnarray}\label{AM-asymptotic-fields}
{\bf J}_{\rm as}&=&\int d^3x \bigg( \psi^\dagger_{\rm as} {\bf x}\times \frac{{\bf \nabla}}{i}\psi_{\rm as}+\psi^\dagger_{\rm as} \frac{{\bf \Sigma}}{2}\psi_{\rm as}+E^i_{\bot {\rm as}} {\bf x}\times {\bf \nabla}A^i_{\bot {\rm as}} \nonumber \\
&&+{\bf E}_{\bot {\rm as}}\times {\bf A}_{\bot {\rm as}} \bigg) \nonumber \\
&=& {\bf L}_{e \rm{as}}+{\bf S}_{e \rm{as}}+{\bf L}_{\gamma \rm{as}}+{\bf S}_{\gamma \rm{as}},
\end{eqnarray}
where the subscript "as" stands for "in" or "out". The operator ${\bf J}_{\rm as}$ thus constructed is the rotation generator for the asymptotic free-particle fields $\psi_{\rm as}$ and ${\bf A}_{\bot {\rm as}}$, and should coincide
exactly with the rotation generator for the interpolating (interacting) fields $\psi$ and ${\bf A}_\bot$.
In the expression (\ref{AM-asymptotic-fields}) a split of ${\bf J}_{\rm as}$ into the OAM and spin contributions of the electron and photon fields is clearly shown. Thus, we have two sets of angular momentum operators, which describe the free electron and photon system at the two asymptotic instants of time $t \rightarrow \pm \infty$.
We also observe that in the separation (\ref{AM-asymptotic-fields}), the individual photon OAM and spin operators are conserved in time: $\frac{d {\bf L}_{\gamma \rm{as}}}{dt}=0,~~\frac{d {\bf S}_{\gamma \rm{as}}}{dt}=0$,
which is a feature of the Coulomb gauge quantization. Now, if one recalls the connection between the asymptotic in-fields and out-fields:
\begin{eqnarray}
\psi_{\rm out}(x) &=& S^{-1} \psi_{\rm in}(x) S \nonumber \\
{\bf A}_{\bot \rm{out}}(x) &=& S^{-1} {\bf A}_{\bot\rm{in}}(x) S,
\end{eqnarray}
where $S$ is the scattering operator of the system, then one immediately obtains the following connection between
the asymptotic photon OAM and spin operators:
\begin{eqnarray}
{\bf L}_{\gamma {\rm out}} =S^{-1} {\bf L}_{\gamma {\rm in}} S \nonumber \\
{\bf S}_{\gamma {\rm out}} =S^{-1} {\bf S}_{\gamma {\rm in}} S.
\end{eqnarray}
From a fundamental point of view, Eq. (\ref{AM-asymptotic-fields}) in fact provides a complete solution to the problem of angular momentum separation of this QED system because in the remote past (or the remote future) any state of the interacting field system is a linear superposition of free-particle in-states (or out-states).
Now, if we try to observe the system at a finite instant $t$, we need to introduce the OAM
and spin operator for the electrons and photons at this specific instant of time. To this end, we
write the total angular momentum operator (the rotation generator) in terms of the interacting fields
at this specific instant $t$:
\begin{eqnarray}\label{AM-interacting-fields}
 {\bf J}&=& \int d^3x \bigg( \psi^\dagger {\bf x}\times \frac{{\bf \nabla}}{i}\psi+\psi^\dagger \frac{{\bf \Sigma}}{2}\psi+E^i_\bot {\bf x}\times {\bf \nabla}A^i_\bot+{\bf E}_\bot\times {\bf A}_\bot \bigg) \nonumber \\
 &=& {\bf L}_{e}(t)+{\bf S}_{e}(t)+{\bf L}_{\gamma}(t)+{\bf S}_{\gamma}(t),
\end{eqnarray}
which represents a natural separation of the conserved angular momentum operator into the OAM and spin contributions
of electron and photon fields at this specific instant. Comparing the separation (\ref{AM-interacting-fields})
with the corresponding asymptotic field separation (\ref{AM-asymptotic-fields}), it is seen that these two separations
have an identical structure, hence in this sense, the finite-time separation (\ref{AM-interacting-fields}) is a natural separation scheme for the angular momentum operator in this interacting QED theory. Then, how are these two
angular momentum separation schemes connected with each other? To understand this point, we recall that in conventional textbooks of quantum field theory one assumes the following formal connection between the interpolating field and its corresponding asymptotic in-field:
\begin{eqnarray}
\psi(x) &=& U^{-1}(t)\psi_{\rm in}(x)U(t) \nonumber \\
{\bf A}_\bot(x) &=& U^{-1}(t){\bf A}_{\bot {\rm in}}(x)U(t) \nonumber \\
{\bf E}_\bot(x) &=& U^{-1}(t){\bf E}_{\bot {\rm in}}(x)U(t),
\end{eqnarray}
where $U(t)$ is a time-dependent unitary operator. Here, we have to remark that due to Haag theorem \cite{Strocchi}, this unitary equivalence between the interacting field and the free asymptotic in-field does not exist in a rigorous mathematical sense. However, in the usual framework of perturbation theory, one accepts the correctness of this statement. If we admit this, we can show that the two angular momentum separation schemes (\ref{AM-interacting-fields}) and (\ref{AM-asymptotic-fields}) are connected by a time-dependent unitary transformation:
\begin{eqnarray}
{\bf L}_{e}(t) &=& U^{-1}(t){\bf L}_{e \rm{in}}U(t) \nonumber \\
{\bf S}_{e}(t) &=& U^{-1}(t){\bf S}_{e \rm{in}}U(t) \nonumber \\
{\bf L}_{\gamma}(t) &=& U^{-1}(t){\bf L}_{\gamma \rm{in}}U(t) \nonumber \\
{\bf S}_{\gamma}(t) &=& U^{-1}(t){\bf S}_{\gamma \rm{in}}U(t).
\end{eqnarray}

With all these constructions at hand, we can say that a formal solution to the angular momentum separation problem
of this interacting QED theory has been found. However, in realistic situations, when stable bound states can be
formed through nontrivial interactions, such a formal solution is inadequate for describing the internal spin structure of this bound state. For example, in a theory where both a phenomenological proton field and an electron field are introduced and interact through their couplings with the electromagnetic field, a stable bound state, i.e.,
the hydrogen atom would appear in the spectrum of the Hamiltonian. If one tries to construct the angular momentum
separation scheme using the asymptotic free-particle fields, then the hydrogen atom
should be introduced as a separate (composite) particle into the whole formalism, with its own asymptotic field.
In this case, one can describe the angular momentum separation for this composite particle as a whole, but cannot use
this angular momentum separation scheme to analyze the internal spin structure of this bound state particle, and the
connection between the angular momentum separation scheme in terms of the interpolating fields and that in terms of the asymptotic fields is much less clear. For the case of non-abelian QCD theory, all the physical particles are colorless hadrons which are bound states of the elementary quark and gluon degrees of freedom, and this problem is
especially apparent. This is just the case for the nucleon spin decomposition problem of QCD.
Within the framework of perturbative QCD, the parton distribution
functions (PDFs) in the nucleon is generally believed to be
quasi-observables, even though they are not genuine observables.
In particular, the so-called gluon spin term in the decomposition
is known to be related to the first moment of the longitudinally
polarized gluon distribution in the nucleon. In the nucleon spin decomposition
problem, a particular direction, i.e., the direction of the nucleon momentum, plays
a special role. It is shown that \cite{Wakamatsu1}, by invoking the physical requirement
of Lorentz-boost invariance along the direction of the parent nucleon momentum,
the uniqueness or nonuniqueness problem of the decomposition of the
total gluon angular momentum has a definite solution.
Once one considers the spin decomposition of a bound state, one encounters
another problem, i.e. the existence of the two types of complete
decomposition of the nucleon spin, the decomposition of the
canonical type and of the mechanical (or kinetic) type. In \cite{Wakamatsu2}
it is shown that what represents the intrinsic spin structure of the nucleon is the
mechanical type decomposition and not the canonical type decomposition.
With all these matters understood, we believe that the investigations we made in this article
can provide some helpful knowledge for the study and understanding of the angular momentum decomposition problem
of gauge field systems.

\section{Acknowledgments}
I thank F. Wang for helpful discussions. This work is supported by the Natural Science Funds of Jiangsu Province of China under Grant No. BK20151376.

\end{document}